# Bloch wavepacket control in truncated modulated optical lattices

Yaroslav V. Kartashov,[1] Victor A. Vysloukh,[2] and Lluis Torner[1]

[1]*ICFO-Institut de Ciencies Fotoniques, and Universitat Politecnica de Catalunya, Mediterranean Technology Park, 08860 Castelldefels (Barcelona), Spain*

[2]*Departamento de Fisica y Matematicas, Universidad de las Americas – Puebla, Santa Catarina Martir, 72820, Puebla, Mexico*

We study the reflection of Bloch wavepackets at the interface of optical lattice possessing a shallow longitudinal out-of-phase refractive index modulation in adjacent waveguides. We show that the relation between the transmitted and reflected energy flows can be efficiently controlled by tuning the frequency and the depth of the modulation. Thus, complete beam reflection may be achieved for a set of resonant modulation frequencies at which light tunneling between adjacent guides of modulated lattice is inhibited.

When a light beam hits the interface between two different linear optical materials it experiences partial reflection and transmission described by Snell's and Fresnel's laws. In the presence of significant nonlinearity, a number of specific effects appear such as hysteresis reflection and optical bistability. A particularly interesting situation arises at interfaces between periodic layered materials. In addition to the fact that such interfaces can support stationary nonlinear surface waves [1,2], even in the linear case a light beam refraction depends not only on the difference of mean refractive indices, but also on the depth and period of refractive index modulation, i.e., it is determined by the band-gap spectrum. This may result in rich dynamics of beams crossing the interface and specific laws of refraction [3-8], as reviewed [9] recently. In addition, a longitudinal refractive index modulation strongly affects the propagation of light in a periodic material. In particular, longitudinal bending of individual waveguides modifies the strength of their coupling, resulting in dynamic localization of light. This effect was first predicted in the simplest two-channel system [10,11] and later observed in periodic lattices [12-15]. Similarly, the coupling between waveguides can be altered by a shallow out-of-phase modulation of refractive index in neighboring guides [16-18] as it was observed in simple lattices [19,20] and suggested in more sophisticated settings [21,22]. Nevertheless, little is known about refraction of light at the boundaries of such longitudinally modulated lattices. Only stationary surface modes have been observed [15,19].

In this Letter we consider the interface between two periodic lattices and suppose that one of them is longitudinally unmodulated, while the refractive index in the neighboring waveguides of other lattice is modulated out-of-phase along the longitudinal direction. We uncover that by tuning the frequency and the depth of shallow longitudinal refractive index modulation one can completely control the Bloch wavepacket reflection at the interface and observe a variety of scenarios ranging from almost complete transmission at modulation frequencies corresponding to strong coupling between the lattice waveguides to total beam reflection for frequencies at which coupling is inhibited.

We describe the propagation of a wave packet in the vicinity of interface of longitudinally modulated and unmodulated lattices by the Schrödinger equation for the dimensionless field amplitude $q$:

$$i\frac{\partial q}{\partial \xi} = -\frac{1}{2}\frac{\partial^2 q}{\partial \eta^2} - pG(\eta,\xi)q. \qquad (1)$$

Here $\eta$ and $\xi$ are the normalized transverse and longitudinal coordinates, respectively; $p$ is the lattice depth proportional to the refractive index contrast; the function $G(\eta,\xi) = G_{\text{left}}(\eta) + G_{\text{right}}(\eta,\xi)$ describes the refractive index distribution in the unmodulated left and modulated right lattices, with $G_{\text{left}} = \sum_{m=1}^{\infty} V(\eta + md)$ and $G_{\text{right}} = \sum_{m=0}^{\infty}[1 + (-1)^m \mu \sin(\Omega_\xi \xi)]V(\eta - md)$; the function $V(\eta) = \exp(-\eta^6/a^6)$ describes transverse refractive index profile of individual guides in both lattices that have the width $a$ and spacing $d$; the parameters $\mu$ and $\Omega_\xi$ stand for the depth and frequency of the longitudinal refractive index modulation. It is supposed that the refractive index in neighboring waveguides in the right lattice is modulated out-of-phase along $\xi$ axis, as it is required for light tunneling inhibition [19]. It should be stressed that the average refractive index is equal at both sides of the interface formed by the lattices. Further we set $a = 0.3$ (waveguides with a width $\sim 3\,\mu m$), $d = 1.5$ (separation $\sim 15\,\mu m$), and lattice depth $p = 12$ (that is equivalent to real refractive index contrast $\delta n \sim 10^{-3}$ at $\lambda = 800\,nm$).

Here we consider the refraction of light beams that fall on the interface from the unmodulated left lattice using the wide Bloch wavepacket as an initial condition in Eq. (1): $q|_{\xi=0} = w(\eta)\exp(ik\eta)\exp[-(\eta - \eta_c)^2/\eta_w^2]$, where the function $w(\eta)$ describes the profile of the Bloch wave corresponding to propagation constant $b(k)$, $k$ is the Bloch momentum, $\eta_{\rm c} = -40d$ and $\eta_{\rm w} = 10$ are the position of center and the width of broad Gaussian envelope. The behavior of the wavepacket inside the left lattice is determined by the dispersion relation $b(k)$ depicted in Fig. 1(a) (we consider only Bloch waves from first band in the lattice spectrum). Thus, for $\eta_{\rm w} \gg d$ such beams move almost without radiation inside the lattice at the angle $-db/dk$ that reaches maximal value around $k = 0.26\Omega_\eta$ (here $\Omega_\eta = 2\pi/d$ is the lattice frequency) [Fig. 1(b)]. The diffractive broadening of beams with $\eta_{\rm w} \gg d$ is negligible on the distances considered here.

When such broad wavepacket hits the interface under consideration the effective coupling between channels in right lattice, which is fully controlled by the depth and frequency of longitudinal modulation (as demonstrated in [19]), determines the fraction of transmitted and reflected energy flows. In order to illustrate the effect of longitudinal modulation on coupling between adjacent guides in the right lattice we plot in Fig. 1(c) the dependence of distance-averaged energy flow $U_{\rm m} = L^{-1} \int_0^L d\xi \int_{-d/2}^{d/2} |q(\eta, \xi)|^2 \, d\eta / \int_{-d/2}^{d/2} |q(\eta, 0)|^2 \, d\eta$ (here $L$ is the propagation distance) trapped inside excited waveguide versus longitudinal modulation frequency $\Omega_\xi$ for a system of only two waveguides with out-of-phase refractive index modulation (further we normalize $\Omega_\xi$ to beating frequency $\Omega_{\rm b}$ in two unmodulated waveguides). One observes that $U_{\rm m} \to 1$ only for a specific resonant modulation frequency $\Omega_\xi = \Omega_{\rm r}$ that indicates on complete light tunneling inhibition. In fact there are multiple secondary resonances at $\Omega_\xi \approx \Omega_{\rm r}/n$, where $n = 1, 2, 3\dots$, but we show only the primary one in Fig. 1(c) (for details see [19]). Since around $\Omega_\xi = \Omega_{\rm r}$ the coupling between channels of right lattice is fully inhibited, one can expect that a broad wavepacket hitting the interface of two lattices will not penetrate into the right lattice and the entire beam will be reflected. As the modulation frequency detunes from $\Omega_{\rm r}$, the coupling between waveguides of the right lattice becomes stronger and one can expect partial wavepacket transmission through the interface.

Figure 2 shows typical refraction scenarios at such interface for different modulation frequencies $\Omega_\xi$. For $\Omega_\xi$ values that are far from primary and any of secondary resonances one observes almost complete transmission of the input beam accompanied, however, by a notable modification of propagation angle despite equality of average refractive indices at both sides of the interface [Fig. 2(a)]. For $\Omega_\xi$ values at which coupling is already partially reduced a considerable fraction of energy flow is reflected at the interface and refraction becomes remarkably stronger [Figs. 2(b) and 2(d)]. Almost complete reflection occurs around $\Omega_\xi = \Omega_{\rm r}$ as well as in all secondary resonances [Fig. 2(c)]. For sufficiently large values of the longitudinal modulation frequency $\Omega_\xi \gg \Omega_{\rm r}$ the refractive index in the right lattice changes too rapidly and has almost no effect on coupling between waveguides, so that one again ob-serves almost complete transmission accompanied by slight refraction [Fig. 2(e)]. Notice that for tilted Gaussian input beams the picture is similar [see Fig. 2(f)].

Figure 3(a) illustrates one of the central results of this Letter - a dramatic dependence of reflection $R = U_{\text{ref}} / U_{\text{in}}$ and transmission $T = U_{\text{tr}} / U_{\text{in}}$ coefficients on the modulation frequency $\Omega_\xi$ (here $U_{\text{ref}} = \int_{-\infty}^{0} |q|^2 \, d\eta$ is the reflected and $U_{\text{tr}} = \int_{0}^{\infty} |q|^2 \, d\eta$ is the transmitted energy flow at $\xi = L$, while $U_{\text{in}} = \int_{-\infty}^{\infty} |q|^2 \, d\eta$ is the input energy flow at $\xi = 0$). One can see that almost complete reflection $(R \to 1)$ occurs in primary and all secondary resonances. As $\Omega_\xi \to \infty$ the transmission coefficient asymptotically approaches unity due to averaging of rapid longitudinal refractive index modulation. The dependencies $R(\Omega_\xi)$, $T(\Omega_\xi)$ for Gaussian inputs are almost identical. Notice that while partial reflection can be observed at the interface of different unmodulated arrays, the increase of waveguides spacing (reduction of coupling) in one of arrays results only in a monotonic decrease of the transmitted energy flow and the angle of refraction, as opposed to the resonant dependencies $R(\Omega_\xi)$, $T(\Omega_\xi)$ at modulated interfaces. Remarkably, such variation of the $R, T$ coefficients is achieved only due to the shallow ($<20\%$) longitudinal refractive index modulation that maintains equal average refractive indexes in both arrays. This is in contrast to the static case, where large modifications of one of the arrays are required. If the focusing nonlinearity is taken into account, the resonances in Fig. 3(a) broaden with increasing peak amplitude, although the amplitude should not exceed the value $\sim 0.2$ to avoid reshaping of Bloch wave-packets in unmodulated array when they propagate toward the interface.

While the magnitude of the angle of reflection $\alpha_{\text{ref}}$ is almost equal to that of the angle of incidence $\alpha_{\text{in}} = -db/dk$, the angle of refraction $\alpha_{\text{tr}}$ strongly and nontrivially depends on the modulation frequency $\Omega_\xi$ [Fig. 3(b)]. It vanishes exactly in the points corresponding to maxima in the $R(\Omega_\xi)$ dependence. When $\Omega_\xi \to \infty$ the refraction angle monotonically increases and approaches $\alpha_{\text{in}}$. Growth of the depth of longitudinal refractive index modulation $\mu$ affects the resonant modulation frequencies at which reflection is strongest. Thus, the resonance frequency $\Omega_r$ increases with $\mu$ almost linearly [Fig. 4(a)], except for the case when $\mu \to 0$ (in this limit the reflection becomes weak and the maximal reflection coefficient $R_{\text{max}}$ rapidly decreases). Interestingly, the smallest width of the reflection band $\delta\Omega_\xi$ (that can be defined as an interval of modulation frequencies around primary resonance within which the reflection coefficient is above $0.9R_{\text{max}}$) is minimal for Bloch momentum $k \approx 0.24\Omega_\eta$. This value is close to $k \approx 0.26\Omega_\eta$ corresponding to largest propagation angle of input beam in unmodulated lattice [Fig. 4(b)]. Detuning of Bloch momentum from $k \approx 0.24\Omega_\eta$ results in broadening of primary reflection band.

Thus, in summary, we introduced a new way to control Bloch wavepackets, which relies on the suitable engineering of the depth and frequency of longitudinal refractive index modulations of periodic optical structures. The effect may be readily observed with currently available material structures.

# References with titles

# References without titles

# Figure captions

Figure 1.    (a) Propagation constant of Bloch wave from first band and (b) derivative $-db/dk$ versus Bloch momentum $k$ in unmodulated lattice. (c) Distance-averaged power fraction trapped in the excited lattice channel versus frequency of longitudinal refractive index modulation at $\mu = 0.2$.

Figure 2.    Dynamics of interaction of input beam with Gaussian envelope whose internal structure corresponds to Bloch wave with $k/\Omega_\eta = 0.26$ with the interface between unmodulated and modulated lattices. In all cases $\mu = 0.2$, while the frequency of longitudinal modulation is (a) $\Omega_\xi = 6.80\Omega_b$, (b) $7.79\Omega_b$, (c) $9.08\Omega_b$, (d) $10.95\Omega_b$, and (e) $16.00\Omega_b$. Panel (f) corresponds to $\Omega_\xi = 9.08\Omega_b$ and simple input Gaussian beam. White dashed lines indicate interface position. The tilt of input beam $\approx 0.13°$ for parameters of our array.

Figure 3.    (a) Reflection $R = U_{ref}/U_{in}$ and transmission $T = U_{tr}/U_{in}$ coefficients versus longitudinal modulation frequency. (b) The propagation angle of the transmitted beam versus longitudinal modulation frequency. In all cases $\mu = 0.2$, $k/\Omega_\eta = 0.26$ [this $k$ value corresponds to circles in Figs. 1(a) and 1(b)].

Figure 4.    (a) Longitudinal modulation frequency corresponding to maximal reflection coefficient $R = R_{max}$ versus $\mu$ at $k/\Omega_\eta = 0.26$. (b) The width of primary reflection band defined at the level $R = 0.9R_{max}$ versus Bloch momentum of input beam at $\mu = 0.2$.

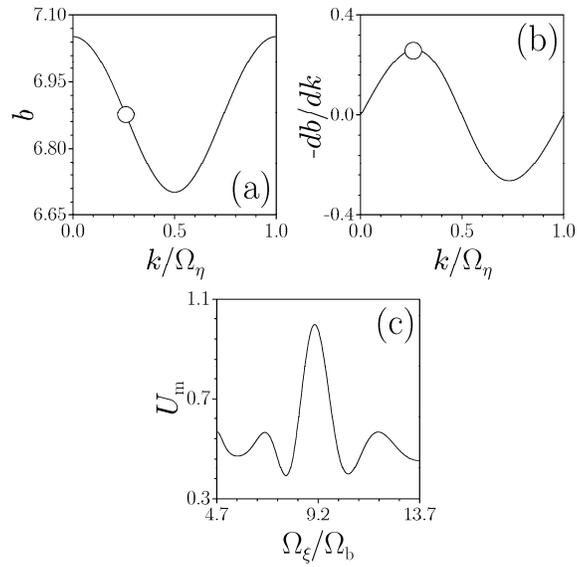

Figure 1.    (a) Propagation constant of Bloch wave from first band and (b) derivative $-db/dk$ versus Bloch momentum $k$ in unmodulated lattice. (c) Distance-averaged power fraction trapped in the excited lattice channel versus frequency of longitudinal refractive index modulation at $\mu = 0.2$.

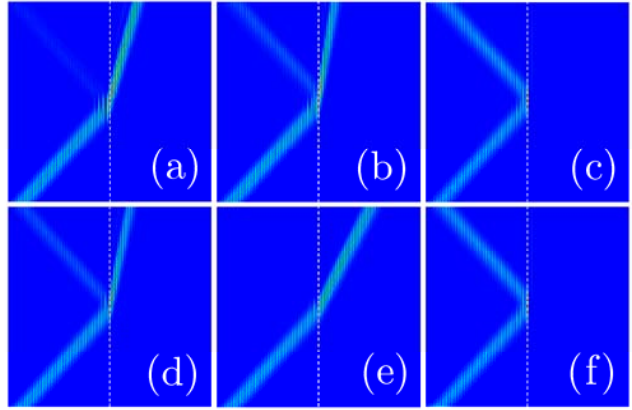

Figure 2.    Dynamics of interaction of input beam with Gaussian envelope whose internal structure corresponds to Bloch wave with $k/\Omega_\eta = 0.26$ with the interface between unmodulated and modulated lattices. In all cases $\mu = 0.2$, while the frequency of longitudinal modulation is (a) $\Omega_\xi = 6.80\Omega_b$, (b) $7.79\Omega_b$, (c) $9.08\Omega_b$, (d) $10.95\Omega_b$, and (e) $16.00\Omega_b$. Panel (f) corresponds to $\Omega_\xi = 9.08\Omega_b$ and simple input Gaussian beam. White dashed lines indicate interface position. The tilt of input beam $\approx 0.13°$ for parameters of our array.

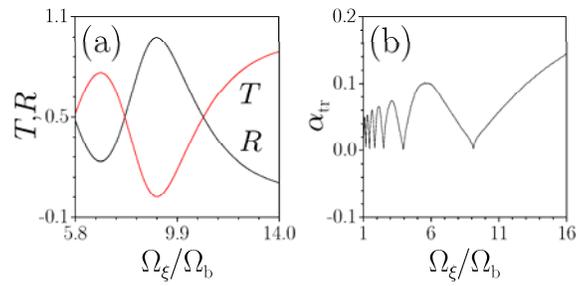

Figure 3.    (a) Reflection $R = U_{\text{ref}} / U_{\text{in}}$ and transmission $T = U_{\text{tr}} / U_{\text{in}}$ coefficients versus longitudinal modulation frequency. (b) The propagation angle of the transmitted beam versus longitudinal modulation frequency. In all cases $\mu = 0.2$ , $k/\Omega_\eta = 0.26$ [this $k$ value corresponds to circles in Figs. 1(a) and 1(b)].

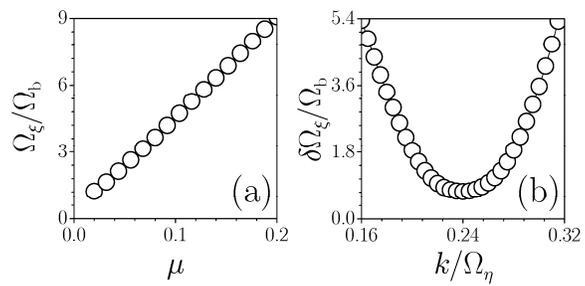

Figure 4.    (a) Longitudinal modulation frequency corresponding to maximal reflection coefficient $R = R_{\max}$ versus $\mu$ at $k/\Omega_\eta = 0.26$. (b) The width of primary reflection band defined at the level $R = 0.9R_{\max}$ versus Bloch momentum of input beam at $\mu = 0.2$.